\documentclass[11pt]{article}

\usepackage[english]{babel}

\usepackage[utf8]{inputenc} % for type accented characters directly
\usepackage{hyperref}
\usepackage{authblk} % for author and affiliation formatting
\usepackage{pdflscape}  % or \usepackage{lscape}
\usepackage{apacite} % for bibliography
\usepackage{siunitx} % for decimal alignment in tables

\usepackage{tablefootnote} % for footnotes in tables

\usepackage{graphicx} % Required for inserting images
\usepackage{booktabs} % For \toprule, \midrule, \bottomrule in tables

\usepackage{soul} % for underline text that can wrap

\usepackage{enumitem} % control indentation of itemized lists

\usepackage{wrapfig} % for wrapping figures
\usepackage{titlesec} % for change spacing between sections
\usepackage{mathpazo}
\usepackage{amsmath}
\usepackage{amssymb}
\usepackage{setspace}
\usepackage{booktabs}
\usepackage{multirow}
\usepackage{booktabs,longtable}
\DeclareMathOperator{\BN}{BN}
\DeclareMathOperator{\LN}{LN}
\DeclareMathOperator{\GELU}{GELU}
\DeclareMathOperator{\Drop}{Drop}
\newcommand{\RoPET}{\mathrm{RoPE}_T}
\newcommand{\RoPEH}{\mathrm{RoPE}_H}
\newcommand{\TrEnc}{\mathrm{TransformerEncoder}}
\newcommand{\vect}[1]{\mathbf{#1}}   % bold for tensors/vectors

\usepackage{algorithm}
\usepackage{algpseudocode}

\usepackage[letterpaper, margin=1in]{geometry}

\usepackage[export]{adjustbox} % for align figures left or right

\titlespacing*{\section}
{0pt}{0.5ex plus 1ex minus .2ex}{0.3ex plus .2ex}
\titlespacing*{\subsection}
{0pt}{0.5ex plus 1ex minus .2ex}{0.3ex plus .2ex}
\titlespacing*{\subsubsection}
{0pt}{0.5ex plus 1ex minus .2ex}{0.3ex plus .2ex}

\begin{document}
\title{Imputing Missing Long-Term Spatiotemporal Multivariate Atmospheric Data with CNN-Transformer Machine Learning}
\author[1]{Jiahui Hu}
\author[1,2,3]{Wenjun Dong\thanks{Corresponding author: \texttt{wenjun@gats-inc.com}}}
\author[1]{Alan Z. Liu}
\affil[1]{Center for Space and Atmospheric Research, Embry-Riddle Aeronautical University, Daytona Beach, FL}
\affil[2]{Global Atmospheric and Science Technologies, inc, Boulder, CO}
\affil[3]{High Altitude Observatory, NSF National Center for Atmospheric Research, Boulder, CO}
\date{} % This hides the date
\maketitle

\section*{Abstract}
Continuous physical domains are important for scientific investigations of dynamical processes in the atmosphere. However, missing data—arising from operational constraints and adverse environmental conditions—pose significant challenges to accurate analysis and modeling. To address this limitation, we propose a novel hybrid Convolutional Neural Network (CNN)–Transformer machine learning model for multivariable atmospheric data imputation, termed CT-MVP. This framework integrates CNNs for local feature extraction with transformers for capturing long-range dependencies across time and altitude. The model is trained and evaluated on a testbed using the Specified Dynamics Whole Atmosphere Community Climate Model with thermosphere and ionosphere extension (SD-WACCM-X) dataset spanning 13 years, which provides continuous global coverage of atmospheric variables, including temperature and zonal and meridional winds. This setup ensures that the ML approach can be rigorously assessed under diverse data-gap conditions. The hybrid framework enables effective reconstruction of missing values in high-dimensional atmospheric datasets, with comparative evaluations against traditional methods and a simple transformer. The results demonstrate that CT-MVP achieves superior performance compared with traditional approaches, particularly in cases involving extended periods of missing data, and slightly outperforms a simple transformer with the same hyper-parameters.

\section*{Plain Language}
Scientists need continuous data to study how the atmosphere behaves, but real-world measurements often contain gaps due to instrument limitations or poor observing conditions. To address the challenge, we developed a hybrid machine learning model called CT-MVP. The model combines two powerful techniques: convolutional neural networks, which capture local patterns, and transformers, which learn long-term trends across time and altitude. We trained and tested CT-MVP on 13 years of global atmospheric data from the state-of-the-art numerical atmospheric model, which provides complete and continuous simulations. This setup allowed us to robustly evaluate the model under diverse conditions. Our results show that CT-MVP outperforms traditional methods, particularly when large sections of data are missing, which indicates machine learning can be a promising tool for reconstructing atmospheric datasets.

\section{Introduction}
The Earth's atmosphere is a highly dynamic system, with interactions between different layers driving key processes that regulate climate and space weather \cite{barry2009atmosphere}. These interactions occur across a wide range of spatial and temporal scales, influencing large-scale circulation, energy transfer, and variability in the upper atmosphere. Continuous atmospheric datasets are essential for enhancing our understanding of coupled dynamical processes across atmospheric layers. However, long-term observational records often contain substantial gaps due to operational faults, sensor malfunctions, adverse environmental conditions, or under-sampling. These missing values pose challenges to the statistical analysis, introducing bias in physical interpretation, and degrading downstream applications such as forecasting and data assimilation. 

Traditional approaches, such as linear interpolation methods that estimate missing values from adjacent time steps, are only effective for imputing short gaps \cite{betancourt2023graph}. More advanced interpolation-based techniques, including kriging and polynomial fitting, remain widely used in environmental science because of their simplicity and interpretability \cite{larson2023reconstructing}. Other Ensemble-based or advanced statistical approaches can outperform classical interpolation when applied to long-term ecological datasets. However, the ensemble methods are built based on smoothness or locality assumptions, and often fail when faced with long missing intervals or nonlinear cross-variable dependencies, where atmospheric processes are strongly coupled and variability spans multiple scales across altitude and time.

Recent advances in Machine Learning (ML) offer alternative solutions to the data gap challenges across a variety of applications \cite{platias2020comparison, emmanuel2021survey,teegavarapu2024imputation}. ML-based imputation methods such as random forests and k-Nearest Neighbors (kNN) have been applied to meteorological datasets, showing greater effectiveness than traditional methods when the proportion of missing data is high \cite{doreswamy2017performance}. Approaches that combine dense layers with convolutional neural network (CNN) have further improved performance by capturing temporal dependencies and spatial patterns across multiple missing variables (e.g., temperature, wind speed, precipitation) measured at stations from the National Climatic Data Center (NCDC). More recently, transformer-based models have demonstrated strong capabilities in capturing long-range dependencies. For instance, \cite{ayub2024enhancing} applied a transformer‑based model to univariate time‑series data aggregated at hourly, daily, and monthly frequencies, achieving significant improvements over classical approaches such as mean and KNN imputation. Innovations such as missing-position encoding \cite{wi2024continuous} and heterogeneous node embeddings have been specifically designed to improve imputation in multivariate, irregularly spaced time series.

By leveraging the capabilities of neural networks to learn both local features and long-range dependencies, advanced techniques that combine different ML architectures can achieve higher accuracy and robustness. The integration of meteorological factors into pollution prediction models has highlighted the value of incorporating domain-specific characteristics to improve imputation performance, as demonstrated by the CNN–LSTM hybrid model for airborne particle forecasting \cite{samal2021improved}. Furthermore, frameworks that integrate convolutional layers with transformer-based architectures have shown superior performance in reconstructing missing air quality datasets \cite{cui2023deep}. Wang et al. \cite{wang2025high} developed a CNN–Transformer model with a customized loss function to predict high-resolution $PM_{2.5}$ from sparse mobile monitoring data, demonstrating the feasibility of fusing local spatial filters with long-range temporal attention for environmental prediction. Similarly, Hou et al. \cite{hou2025hybrid} proposed a CNN–Transformer model to interpolate missing meteorological variables on the Tibetan Plateau, reporting that the hybrid design significantly outperformed conventional machine learning and statistical interpolation methods. These studies underscore the growing recognition that hybrid architectures can balance fine-scale feature extraction with long-range dependency modeling. However, most prior work has focused on either urban pollutant mapping or single-site meteorological interpolation, often with limited variables or restricted spatiotemporal coverage.

Inspired by recent ML methods for meteorological data imputation, we develop a hybrid CNN–Transformer framework tailored to the time–altitude domain. The model combines convolutional encoders for local feature extraction with transformer layers enhanced by rotary embeddings \cite{su2024roformer}, which is expected to improve the capture of long-range temporal–vertical dependencies. Our task is distinct as it targets multivariate atmospheric data (e.g., temperature, zonal and meridional wind), where gaps impact not only statistical fidelity but also the physical interpretability of wave propagation and mesospheric tides. This design ensures that reconstructions are not only numerically accurate but also scientifically meaningful to preserve gradients, oscillations, and variability that underpin physical coupling across atmospheric layers.

Because spatiotemporal observations for training are not sufficient to assess model performance, we use the Specified Dynamics Whole Atmosphere Community Climate Model with thermosphere and ionosphere extension (SD-WACCM-X) as a controlled testbed, where artificial gaps are introduced to mimic realistic observational gaps of varying duration. This setup allows us to robustly test model skill in reconstructing missing multi-variate atmospheric data. We benchmark CT-MVP against widely used linear interpolation method and advanced statistical imputation approaches, including Rauch–Tung–Striebel (RTS) Kalman filtering and smoothing \cite{sarkka2008unscented} and Principal Component Analysis (PCA) \cite{abdi2010principal}, as well as a simple transformer \cite{vaswani2017attention}. The results demonstrate that CT-MVP achieves higher reconstruction accuracy than traditional methods, slightly better than a simple transformer, particularly in extended-gap cases. These findings highlight the potential of machine learning frameworks for atmospheric data imputation, with strong prospects for application to real-world observational records.

\section{CNN-Transformer Multi-Variable imPutation}
\label{sect: Method}
We propose CT-MVP, a hybrid ML approach specifically designed for time–altitude data imputation, addressing the challenge of missing values across multiple physical variables (e.g., zonal and meridional neutral winds, temperature). The details of the CT-MVP architecture are presented in Subsect. \ref{subsect: Model Architecture}, including the mathematical formulations of the neural network layers and the model flowchart. The experimental setup used to evaluate CT-MVP, validating against existing traditional methods such as linear interpolation, RTS Kalman filtering and smoothing, PCA, and a simple transformer—is described in Subsect. \ref{subsect: Training, Validation, and Test}.

\subsection{Model Architecture}
\label{subsect: Model Architecture}
The first stage of the model employs a CNN encoder consisting of three convolutional layers, each followed by batch normalization (BN) and a ReLU activation. This block captures fine-grained spatiotemporal features across the time and altitude dimensions. The channel dimension (variables) is progressively projected from $d_v$ to $d_m/2$, and finally to the full embedding size $d_m$.

\begin{align}
\vect{Z}_1 &= \mathrm{ReLU}\!\big(\BN_1(\vect{W}_1 \ast \vect{X})\big) \\
\vect{Z}_2 &= \mathrm{ReLU}\!\big(\BN_2(\vect{W}_2 \ast \vect{Z}_1)\big) \\
\vect{X}_{\mathrm{CNN}} &= \mathrm{ReLU}\!\big(\BN_3(\vect{W}_3 \ast \vect{Z}_2)\big)
\end{align}
Where \[
\mathbf{X} \in \mathbb{R}^{d_b \times d_T \times d_h \times d_v},
\qquad
\mathbf{X}_{\mathrm{CNN}} \in \mathbb{R}^{d_b \times d_T \times d_h \times d_m}
\]
\[
\mathbf{W}_1 \in \mathbb{R}^{\frac{d_m}{4} \times d_v \times 3 \times 3},\qquad
\mathbf{W}_2 \in \mathbb{R}^{\frac{d_m}{2} \times \frac{d_m}{4} \times 3 \times 3},\qquad
\mathbf{W}_3 \in \mathbb{R}^{d_m \times \frac{d_m}{2} \times 3 \times 3}.
\]
\[
\mathbf{Z}_1 \in \mathbb{R}^{d_b \times d_T \times d_h \times \frac{d_m}{4}},\qquad
\mathbf{Z}_2 \in \mathbb{R}^{d_b \times d_T \times d_h \times \frac{d_m}{2}},\qquad
\]

After extracting spatial features, a channel mixer implemented as a two-layer multilayer perceptron (MLP) with Gaussian Error Linear Unit (GELU) nonlinear activation, dropout that mixes information across variables, as well as a residual connection and Layer Normalization (LN) stabilize training:

\begin{align}
\widetilde{\vect{X}} &= \vect{W}_4\,\Drop\!\Big(\GELU\!\big(\vect{W}_3 \vect{X}_{\mathrm{CNN}} + \vect{b}_3\big)\Big) + \vect{b}_4 \\
\vect{X}_{\mathrm{mix}} &= \LN\!\big(\vect{X}_{\mathrm{CNN}} + \widetilde{\vect{X}}\big)
\end{align}
Where \[
\widetilde{\vect{X}} \in \mathbb{R}^{d_b \times d_T \times d_h \times d_m}, \qquad
\vect{X}_{\mathrm{mix}} \in \mathbb{R}^{d_b \times d_T \times d_h \times d_m}
\]
\[
\vect{W}_3 \in \mathbb{R}^{2d_m\times d_m}, \qquad \vect{W}_4 \in \mathbb{R}^{d_m\times 2d_m}, \qquad 
\vect{b}_3 \in \mathbb{R}^{2d_m}, \qquad
\vect{b}_4 \in \mathbb{R}^{d_m}
\]

We then apply rotary positional embeddings along time and altitude to encode relative positions. Denoting $\mathrm{RoPE_T}$ and $\mathrm{RoPE_H}$ as the rotary maps applied along $t$ and $h$, respectively. The output of $\vect{X}_{\mathrm{rope}}$ has the same dimension of $\vect{X}_{\mathrm{mix}}$.

\begin{equation}
\vect{X}_{\mathrm{rope}} \;=\; \RoPEH\!\big(\RoPET(\vect{X}_{\mathrm{mix}})\big)
\end{equation}

The transformer encoder operates on the flattened $(t,h)$ grid ($d_Td_H$ tokens per batch item), capturing long-range dependencies within each window via multi-head self-attention (pre-norm, GELU feed-forward):

\begin{align}
\vect{X}_{\mathrm{tr}} &= \TrEnc\!(X_{\mathrm{flat}})
\end{align}
Where \[
\vect{X}_{\mathrm{flat}} \in \mathbb{R}^{d_b \times (d_Td_h) \times d_m}, \qquad
\vect{X}_{\mathrm{tr}} \in \mathbb{R}^{d_b \times d_T \times d_h \times d_m}
\]

Finally, a two-layer output head projects back to the original variable space to produce imputed values:

\begin{equation}
\vect{X}_{\mathrm{pred}} \;=\; 
\vect{W}^{(2)}_{\mathrm{out}}\,
\Drop\!\Big(\GELU\!\big(\vect{W}^{(1)}_{\mathrm{out}} (\vect{X}_{\mathrm{tr}}+\vect{X}_{\mathrm{rope}})\big)\Big)
+ \vect{b}^{(2)}_{\mathrm{out}}
\end{equation}
Where
\[
\vect{W}^{(1)}_{\mathrm{out}} \in \mathbb{R}^{\frac{d_m}{2} \times d_m}, \qquad \vect{b}^{(1)}_\mathrm{out} \in 
\mathbb{R}^{\frac{d_m}{2}}, \qquad
\vect{W}^{(2)}_{\mathrm{out}} \in \mathbb{R}^{d_v \times \frac{d_m}{2}}, \qquad \vect{b}^{(2)}_\mathrm{out} \in 
\mathbb{R}^{d_v}
\]

\begin{figure}[!b]
    \centering
    \includegraphics[width=\textwidth]{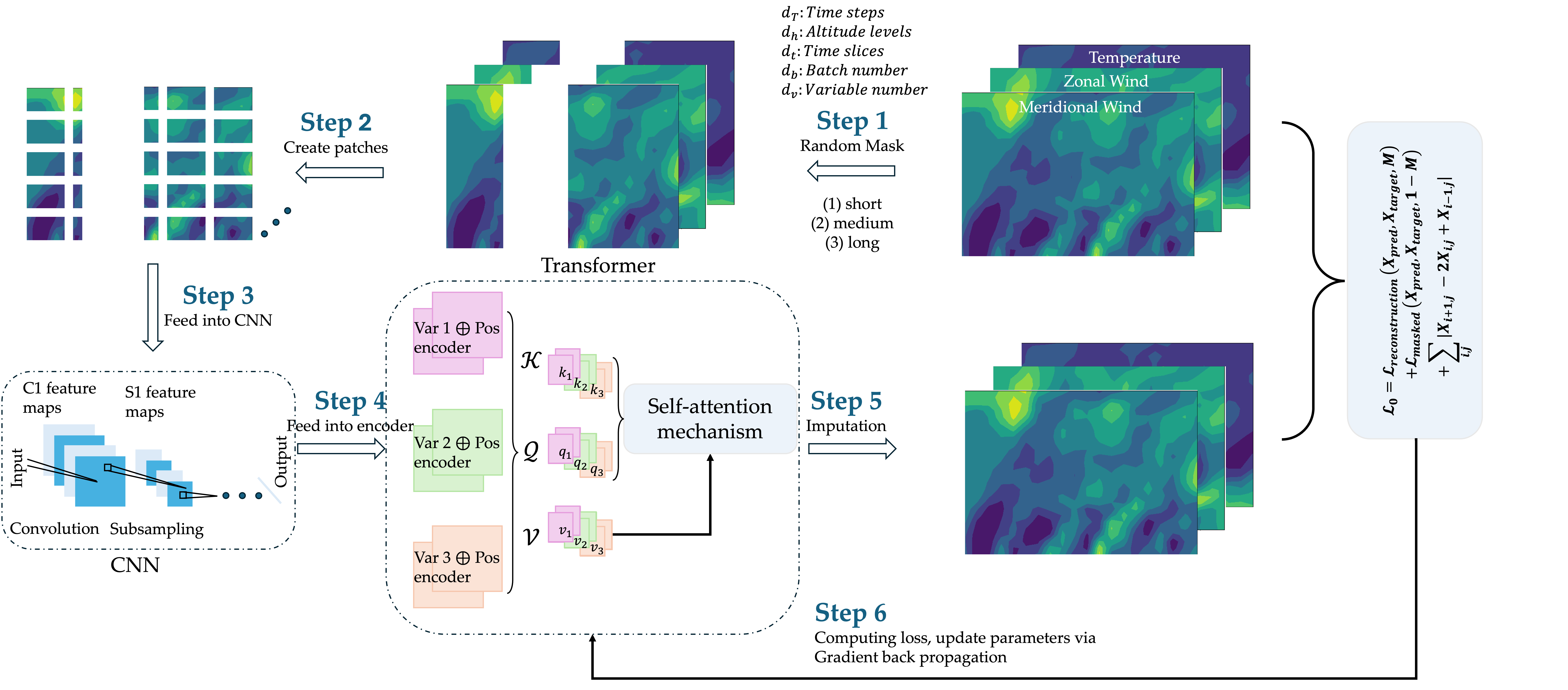}
    \caption{Schematic of the multivariable masked transformer used for atmospheric data imputation. Temperature, zonal wind, and meridional wind fields are masked, patch-encoded with positional information, and processed through a self-attention mechanism. The model reconstructs complete fields by minimizing a composite loss combining reconstruction, masked fidelity, and smoothness constraints.}
    \label{fig: CT-MVP flowchart}
\end{figure}

Fig. \ref{fig: CT-MVP flowchart} illustrates the workflow of the CNN–Transformer framework for imputing atmospheric multivariate data. In Step 1, temperature, zonal and meridional wind are extracted as time–altitude profiles, and random masking is applied to simulate missing observations, leaving blank regions for the model to reconstruct. In Step 2, the masked inputs are divided into smaller spatiotemporal patches. In Step 3, these patches are processed through convolutional encoders to capture localized features. Step 4 embeds each variable with rotary positional information and projects them into latent representations. Step 5 employs a self-attention mechanism to capture both cross-variable dependencies and long-range correlations across time and altitude. Finally, in Step 6, the model reconstructs the complete multivariate fields by minimizing a composite loss that combines reconstruction error, masked region error, and smoothness regularization.

Tab. \ref{tbl: hyper-parameters} lists the hyperparameters used in both CT-MVP and a simple transformer. Both architectures share a common Transformer backbone with an embedding dimension of 128, 8 attention heads, 6 encoder layers, and a feedforward hidden size of 512, with dropout set to 0.1. The key differences lie in their input encoding and positional representations. The CT-MVP model employs a convolutional encoder with three successive Conv2d–BatchNorm–ReLU blocks that expand channels from 3 to 128, followed by a channel mixer MLP ($128 \rightarrow 256 \rightarrow 128$) with residual connections and LayerNorm. Positional information is incorporated using rotary embeddings applied independently along the time and altitude dimensions, each with a per-head rotary dimension of 16. In contrast, the Simple Transformer omits convolutional preprocessing and channel mixing, instead applying a linear projection from $3 \rightarrow 128$ and learned embedding layers for time and altitude are 256. Both models conclude with a Transformer encoder stack and an output head MLP mapping $128 \rightarrow 64 \rightarrow 3$ with GELU activation and dropout, followed by a final LayerNorm dimension of 128. This design ensures a fair comparison between a lightweight pure Transformer as one of the baselines and a CNN-integrated transformer with rotary embeddings.

\begin{table}[!h]
\centering
\caption{Model hyper-parameters for the two ML architectures.}
\begin{tabular}{p{4cm} p{5cm} p{5cm}}
\toprule
\textbf{Hyper-parameter} & \textbf{CT-MVP} & \textbf{Simple Transformer} \\
\midrule
Input channels & 3 (T, U, V) & 3 (T, U, V) \\
Embedding dimension & 128 & 128 \\
Attention heads     & 8 & 8 \\
Encoder layers                 & 6 & 6 \\
Feedforward hidden size & 512 & 512 \\
Dropout                        & 0.10 & 0.10 \\
\midrule
CNN encoder                    & 3×Conv2d + BN + ReLU & -- \\
CNN channels                   & $3 \rightarrow 32 \rightarrow 64 \rightarrow 128$ & -- \\
Channel mixer           & $128 \rightarrow 256 \rightarrow 128$ + Res + LN & -- \\
Input projection      & -- & $3 \rightarrow 128$ \\
\midrule
Positional encoding (time \& altitude)     & Rotary per-head dim=16 & Learned embedding=256 \\
\midrule
Transformer norm style         & Pre-norm, final LayerNorm(128) & Pre-norm, final LayerNorm(128) \\
Output head (MLP)              & $128 \rightarrow 64 \rightarrow 3$ (GELU+Dropout) & $128 \rightarrow 64 \rightarrow 3$ (GELU+Dropout) \\
\bottomrule
\label{tbl: hyper-parameters}
\end{tabular}
\end{table}

\subsection{Training, Validation, and Test}
\label{subsect: Training, Validation, and Test}
The proposed CT-MVP model is trained and evaluated using multi-variables from SD-WACCM-X, which provides continuous global atmospheric fields at six-hourly resolution. The data is windowed into fixed-length segments of 20 time epochs as a data batch, with each sample window represented as a tensor of size $[d_t \times d_h \times d_v]$, where $d_t$ denotes the temporal length, $d_h$ as the number of altitude levels, and $d_v$ as the number of variables. To emulate observational data gaps, artificial data gaps is introduced by randomly masking contiguous time intervals of varying duration. The masked windows are used as model inputs, and the corresponding unmasked data serves as ground-truth targets.

The dataset is divided by calendar year into training (2000–2010), validation (2011–2012), and test (2013) sets. Temporal–vertical profiles are extracted from five mountain regions across different continents: North America (Rocky Mountains, $39.6^{\circ}\mathrm{N},106.4^{\circ}\mathrm{W}$), South America (Andes, $32.7^{\circ}\mathrm{S},70^{\circ}\mathrm{W}$), Europe (Alps, $45.8^{\circ}\mathrm{N},6.9^{\circ}\mathrm{E}$), Asia (Tien Shan, $42.3^{\circ}\mathrm{N},78.3^{\circ}\mathrm{E}$), and Africa (Atlas Mountains, $31^{\circ}\mathrm{N},7.9^{\circ}\mathrm{W}$), with grid points chosen closest to the actual mountain locations. Two validation sites are selected near North America ($46.85^{\circ}\mathrm{N},121.87^{\circ}\mathrm{W}$) and Asia ($35.36^{\circ}\mathrm{N},138.72^{\circ}\mathrm{E}$).

To rigorously assess imputation performance, we define three gap scenarios that reflect common patterns of data loss in atmospheric observations. In the short-gap scenario, $20\%$ of the time steps are randomly masked in contiguous blocks of one day (equivalent to 4 epochs), simulating temporary outages such as brief sensor malfunctions or transmission interruptions. The medium-gap scenario masks $40\%$ of the time steps in two-day blocks, representing more sustained data gaps that could arise from adverse environmental conditions or multi-day instrument downtime. Finally, the long-gap scenario masks up to $60\%$ of the time steps in three-day blocks, mimicking extended data losses similar to prolonged observational gaps in ground-based campaigns. Together, these scenarios span a range of realistic missing-data conditions, from short outages to extended gaps, enabling a systematic evaluation of CT-MVP’s robustness compared with other imputation techniques.

For benchmarking, the same masked test sets are reconstructed using a set of traditional imputation methods, including linear interpolation, PCA, and RTS Kalman filtering and smoothing. These approaches were chosen because they represent widely used strategies in environmental and atmospheric sciences, spanning from simple statistical fillers to more advanced model-based techniques. Linear interpolation is a common choice for filling short-term observational outages, as it enforces temporal continuity but is known to degrade under long or nonlinear variability. PCA exploits the low-rank structure of multivariate datasets by projecting incomplete data onto a reduced set of dominant modes, making it effective when variability is governed by a few principal components, but limited in capturing localized or nonlinear dynamics. RTS Kalman filtering and smoothing applies a state-space formulation with recursive updates to estimate missing values, incorporating temporal correlation and uncertainty propagation, but relies on the restrictive assumption of linear-Gaussian dynamics. Together, these traditional approaches provide a spectrum of baseline performance levels, allowing us to highlight not only the gains achieved by CT-MVP but also the types of atmospheric structures that simpler methods tend to miss.

Performance is evaluated using the error metrics of mean squared error (MSE) and mean absolute error (MAE) to quantify amplitude differences, as well as the Pearson correlation coefficient (R) to assess temporal and vertical pattern fidelity, and total relative variation difference ($\Delta TV$ \%) to measure structural consistency in the reconstructed fields. Metrics are reported both as averages across all data batches and locations, as variable-specific results to capture differences in dynamical behavior among temperature and winds. Beyond the metrics, we also present qualitative diagnostics that visualize the ground truth, imposed gaps, reconstructed values, and associated absolute errors for representative samples. This combination of quantitative and visual evaluation provides a multi-scale assessment, ensuring that CT-MVP is judged not only by numerical accuracy but also by its ability to reproduce physically meaningful structures when compared against other imputation methods.

\section{Results}
\label{sect: Results}
\begin{figure}
    \centering
    \includegraphics[width=0.8\linewidth]{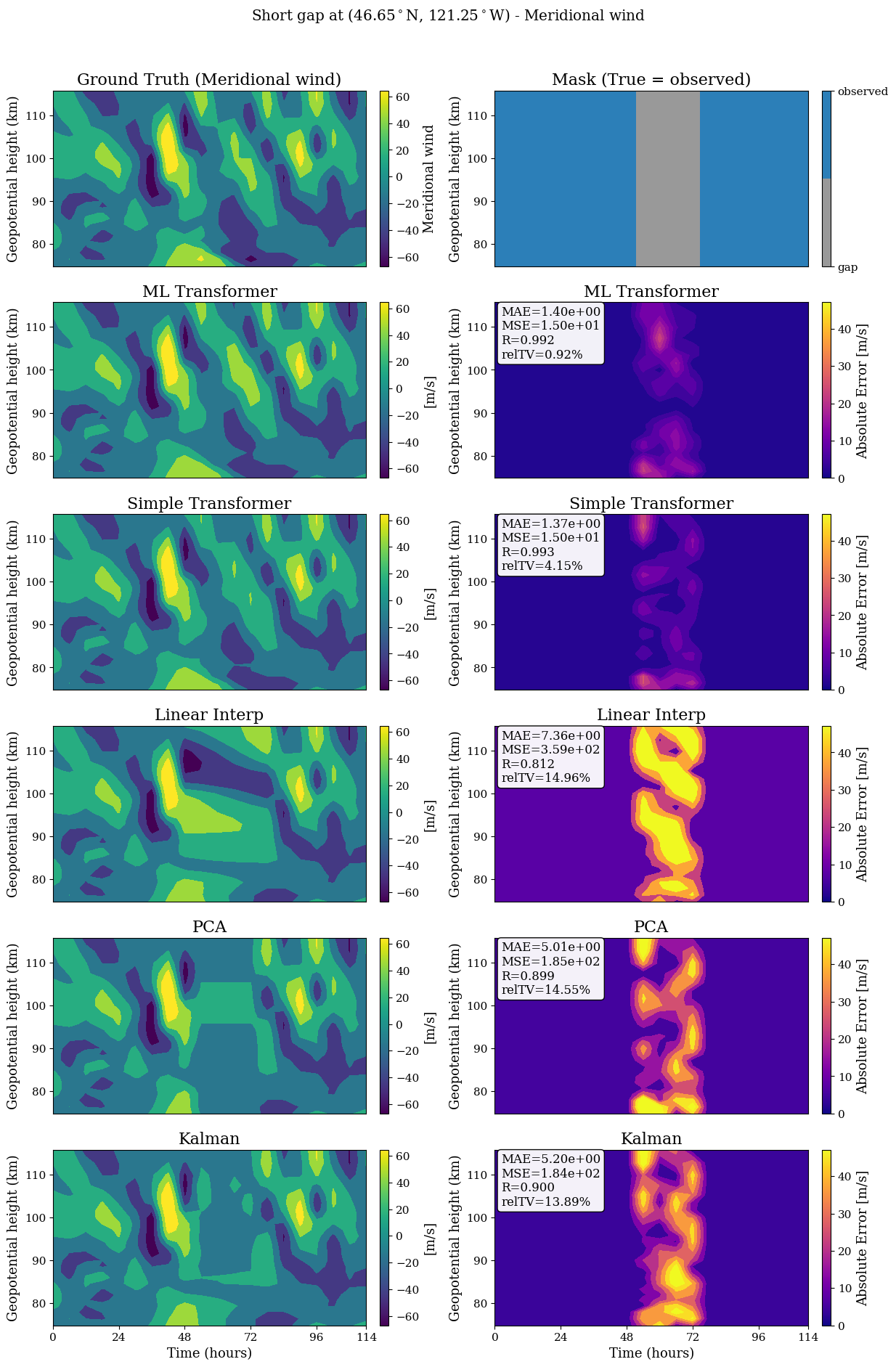}
    \caption{Example gap–filling comparison for the meridional wind ($V$) near $46.65^{\circ}\mathrm{N}$, $121.25^{\circ}\mathrm{W}$. Left column shows the ground truth (top) and each method’s reconstruction; right column shows the mask (blue = observed, gray = gap) and the absolute error. All reconstructions share a common color scale (m\,s$^{-1}$); all error maps share a common absolute-error scale (m\,s$^{-1}$). Time is in 6\,h steps (x-axis), altitude is geopotential height (km, y-axis). Inset boxes report full-field MAE, MSE, Pearson correlation $R$, and relative total variation error ($\Delta \mathrm{TV}_r$ \%).}
    \label{fig:batch comparison - short gap}
\end{figure}
%\begin{figure}
%    \centering
    %\includegraphics[width=\linewidth]{figures/sample000_medium_gaps_gaps_loc0_V.png}
    %\caption{Caption}
    %\label{fig:batch comparison - medium gap}
%\end{figure}
\begin{figure}
    \centering
    \includegraphics[width=0.8\linewidth]{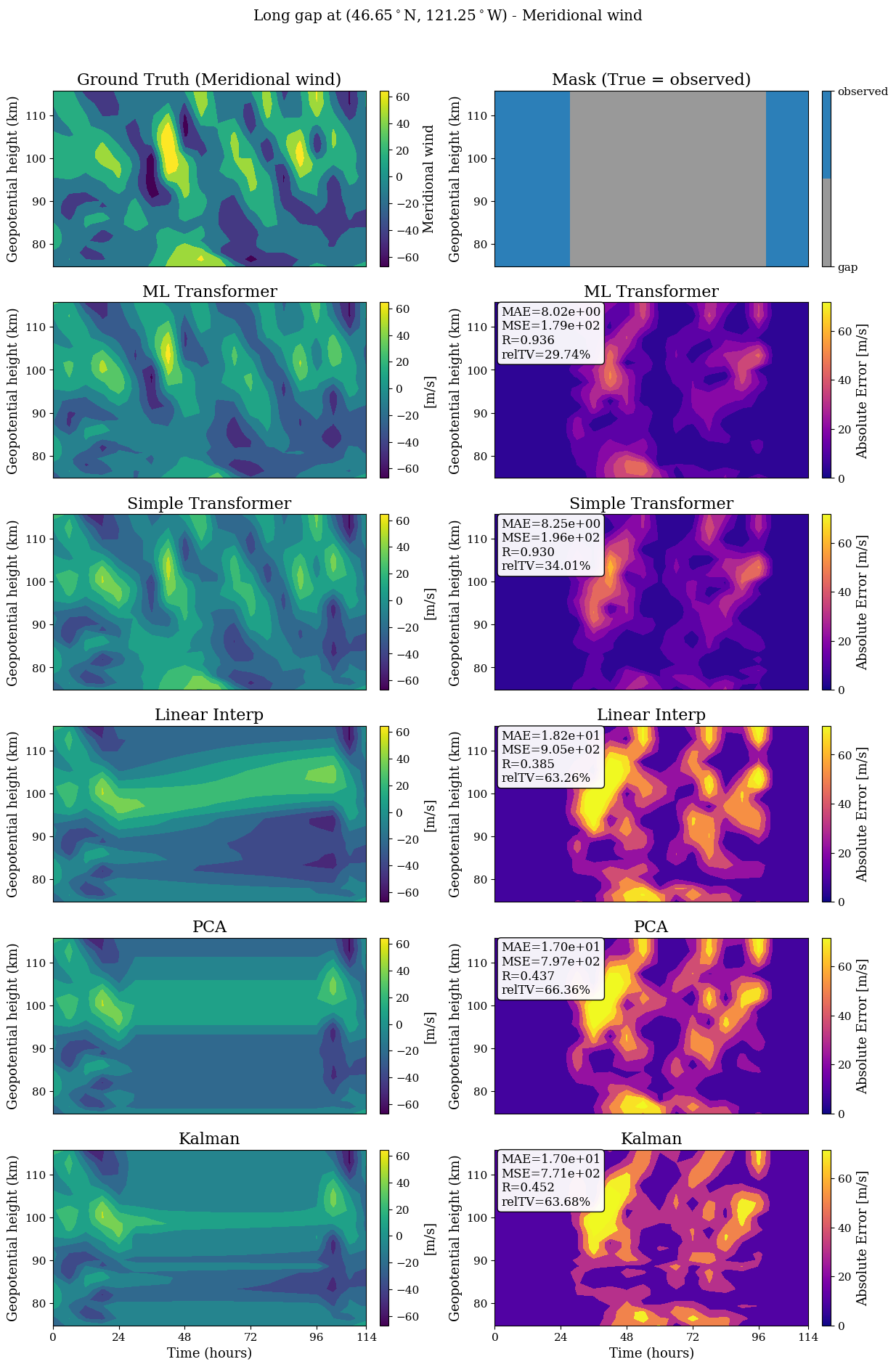}
    \caption{Same as Fig.~\ref{fig:batch comparison - short gap}, but for a more challenging long gap duration centered in time. }
    \label{fig:batch comparison -long gap}
\end{figure}
Across both case studies of filling short and long gaps, Figs. \ref{fig:batch comparison - short gap} and \ref{fig:batch comparison -long gap} shows the ML-based methods (both CT-MVP and simple transformer) clearly outperform the traditional baselines and better preserve the physical structure of the flow. 

In the short-gap example at $46.65^\circ N, 121.25^\circ W$ (6-hour cadence; 75-115 km), the two ML variants recover near-truth meridional wind profiles, with MAE $\approx 1.5 ms^{-1}$, R $\approx 0.99$, and small relative total variation error of $0.92 \%$ for CT-MVP, $4.15\%$ for a simple transformer. In contrast, the traditional methods yield large error in the missing interval and smear vertical gradients. The linear interpolation method yields MAE of $7.36 ms^{-1}$ with R = 0.81, while PCA and KFS give MAE of 5.01 and 5.20 $ms^-1$ respectively, with R = 0.9, and relative $\Delta TV \approx 14 \%$.

With a longer gap, errors grow for all methods, but the machine learning advantage persists. The ML models keep MAE $\approx 8 ms^{-1}$, R $\approx 0.93$, and $\Delta TV_r = 29\%$ for CT-MVP, $25 \%$ for a simple transformer, whereas linear interpolation yields MAE of $18.2 ms^{-1}$ with R = 0.38, and PCA/Kalman filtering-smoothing method yields MAE of $17 ms^{-1}$ with R $\approx 0.45$ and $\Delta TV_r \approx 66\%$. Visually, the traditional methods over-smooth or distort fine scale vertical structure across 80-115 $km$, while ML reconstructions retain sharper shears. 

\begin{figure}
    \centering
    \includegraphics[width=\linewidth]{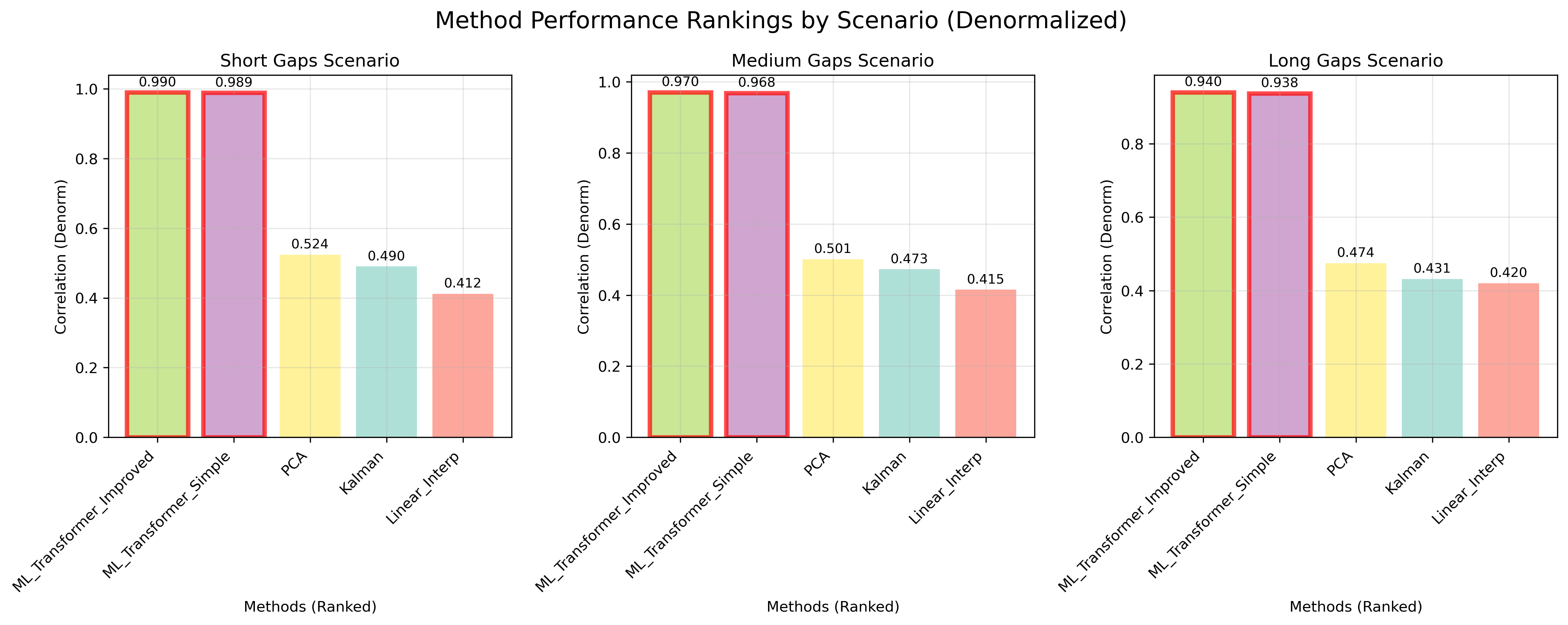}
    \caption{Method performance by gap length. Each panel shows the average correlation between reconstructions and truth for gaps, ranked left-to-right by score, for short, medium, and long gaps scenarios. Bars with a red outline are the two transformer models. }
    \label{fig:corr_comparison}
\end{figure}

Figure \ref{fig:corr_comparison} shows the averaged correlations for the masked regions, and the panels rank imputation methods by denormalized correlation for short, medium and long gap scenarios. In every case, CT-MVP and the simple transformer achieve averaged correlations of 0.99 for short gaps, 0.97 for medium gaps, and 0.94 for long gaps, retaining high skills as gaps increase. Traditional baselines remain well behind and degrade slightly as gap length increases, PCA drops from 0.52 to 0.47, Kalman from 0.49 to 0.43, while linear interpolation remains to be the worst method of maintaining the structural fidelity ($R\approx0.4$). For comprehensive and complete comparison, the full table \ref{tbl: error metrics} in Appendix listed all the averaged error metrics over the multi-variables across different missing data scenarios.

On the same machine, inference for the two ML-based models completed in about 12.0 to 12.5 seconds per scenario, whereas the Kalman filtering and smoothing required 17.5 seconds, PCA required 647 seconds due to more iterations, and linear interpolation was essentially free (0.26 seconds). Thus, the transformer delivers state-of-the-art accuracy at approximately 52 to 54 $\times$ faster than the PCA, and are about 1.4 $\times$ faster than Kalman. The improved transformer is only 3.8 \% slower than the simple variant.

\section{Conclusion and Discussion}
\label{sect: Discussion and Summary}
In this study, we performed synthetic data imputation tasks using different methods, which includes the hybrid CNN-transformer (CT-MVP), a simple transformer, PCA, RTS Kalman filtering and smoothing and linear interpolation. The comparative analysis highlights the advantages of the CT-MVP model across different gap lengths. 

The results that ML approach outperforms other conventional methods demonstrate that machine learning-based imputation is not only feasible but also advantageous for atmospheric applications where extended gaps are common. Kalman filtering and smoothing assume a linear state-space model with Gaussian noise, making the smoothed estimate a fixed linear function of available measurements. While optimal under those assumptions, they fail to capture the strongly nonlinear, non-Gaussian, and multi-variable couplings of atmospheric dynamics, leading smoothed estimations. PCA imputes missing values by projecting onto a low-rank subspace, capturing dominant variability but missing nonlinear or regime-dependent patterns, while linear interpolation simply connects neighboring points, preserving short-term continuity but breaking down for rapid transitions. Together, these methods are computationally simple but structurally restrictive, motivating the need for more flexible ML approaches.

Surprisingly, the CNN-transformer and simple transformer perform almost identically. We think it's because the gap-filling task is dominated by local and smooth structure over a small window (20 time epochs x 26 altitudinal levels). In that regime, both architectures have enough capacity to capture the nearby time-altitude patterns that drive imputation, so the two ML variants yield similar performance. 

Because observational datasets often suffer from weather-related issues, operational outages, or scheduling constraints, they naturally contain various gaps, making it difficult to conduct fair comparison tests across different imputation techniques. The importance of controlled experiments using WACCM-X datasets is to indicate that the ML framework holds promise for real observational datasets such as LiDAR, Radar, and satellite measurements, if the observational datasets are adequate for training. After comprehensively collecting all available measurements, future work should focus on adapting CT-MVP for heterogeneous observational inputs, then integrating physical constraints, such as mass conservation, can further improve reliability and explainability of ML-based methods \cite{urco2024augmented,karniadakis2021physics}. 

In summary, CT-MVP provides a scalable machine learning approach for reconstructing spatiotemporal atmospheric multivariate datasets. its superior performance especially under long-gap conditions makes it a valuable tool for atmospheric science community, enabling more continuous records for climate research, model evaluation, and data assimilation applications.
\appendix
\section{Mathematical formulations of traditional methods}
We denote the true field by $Y_{t,h,v} \in \mathbb{R}^{d_t \times d_h \times d_v}$ 
and the reconstructed field by $\hat{Y}_{t,h,v}$. Observations are available only when the binary mask $M_{t,h,v}=1$; 
otherwise values are missing.

\subsection*{A.1 Linear Interpolation}
For each height--variable column $(h,v)$, consider the time series $\{Y_t\}$ with mask $\{M_t\}$. 
For missing time index $t$ lying between two observed times $t_i < t < t_{i+1}$, interpolation is
\begin{equation}
\hat{Y}_t =
\begin{cases}
Y_{t_1}, & t < t_1, \\[6pt]
Y_{t_i} + \dfrac{Y_{t_{i+1}} - Y_{t_i}}{t_{i+1}-t_i}\,(t - t_i), & t_i \le t \le t_{i+1}, \\[12pt]
Y_{t_K}, & t > t_K ,
\end{cases}
\end{equation}
where $\{t_k\}_{k=1}^K$ are the observed indices.

\subsection*{A.2 Rauch--Tung--Striebel Kalman Filtering and Smoothing}
Each $(h,v)$ column is modeled as a scalar linear Gaussian state--space system:
\begin{align}
Y_t &= a Y_{t-1} + \omega_t, & \omega_t \sim \mathcal{N}(0,q), \\
Z_t &= Y_t + \varepsilon_t, & \varepsilon_t \sim \mathcal{N}(0,r),
\end{align}
with prior $Y_0 \sim \mathcal{N}(\mu_0,P_0)$. 
Let $\hat{Y}_{t|s}$ and $P_{t|s}$ denote the conditional mean and variance.

For the prediction step, 
\begin{align}
\hat{Y}_{t|t-1} &= a \hat{Y}_{t-1|t-1}, \\
P_{t|t-1} &= a^2 P_{t-1|t-1} + q.
\end{align}

The innovation and variance is defined as:
\begin{align}
\nu_t &= Z_t - \hat{Y}_{t|t-1}, \\
S_t &= P_{t|t-1} + r.
\end{align}

The Kalman gain is calculated as:
\begin{equation}
K_t =
\begin{cases}
\dfrac{P_{t|t-1}}{S_t}, & M_t=1, \\
0, & M_t=0.
\end{cases}
\end{equation}

To update the state:
\begin{align}
\hat{Y}_{t|t} &= \hat{Y}_{t|t-1} + K_t \nu_t, \\
P_{t|t} &= (1-K_t) P_{t|t-1}.
\end{align}
If $M_t=0$, then $\hat{Y}_{t|t}=\hat{Y}_{t|t-1}$ and $P_{t|t}=P_{t|t-1}$.

For the Kalman smoothing step, initialize with terminal values $\hat{Y}_{T|T}, P_{T|T}$. 
For $t=T-1,\dots,1$, define smoother gain
\begin{equation}
J_t = \frac{P_{t|t} \, a}{P_{t+1|t}},
\end{equation}
and compute
\begin{align}
\hat{Y}_{t|T} &= \hat{Y}_{t|t} + J_t(\hat{Y}_{t+1|T} - \hat{Y}_{t+1|t}), \\
P_{t|T} &= P_{t|t} + J_t(P_{t+1|T} - P_{t+1|t})J_t.
\end{align}

The actual imputation step is defined as:
\begin{equation}
\hat{Y}_{t,h,v} =
\begin{cases}
Z_{t,h,v}, & M_{t,h,v}=1, \\
\hat{Y}_{t,h,v|T}, & M_{t,h,v}=0.
\end{cases}
\end{equation}

To update the a,q,r recursively, With $P_{t,t-1|T}=\text{Cov}(Y_t,Y_{t-1}|Z_{1:T})$,
\begin{align}
a &\leftarrow \frac{\sum_{t=2}^T \big(P_{t,t-1|T}+\hat{Y}_{t|T}\hat{Y}_{t-1|T}\big)}{\sum_{t=2}^T \big(P_{t-1|T}+\hat{Y}_{t-1|T}^2\big)}, \\
q &\leftarrow \frac{1}{T-1}\sum_{t=2}^T \big(P_{t|T} + a^2 P_{t-1|T} - 2a P_{t,t-1|T} + (\hat{Y}_{t|T}-a\hat{Y}_{t-1|T})^2\big), \\
r &\leftarrow \frac{1}{|\Omega|}\sum_{t \in \Omega} \big((Z_t-\hat{Y}_{t|T})^2 + P_{t|T}\big).
\end{align}

\subsection*{A.3 Principal Component Analysis (PCA)}
For each variable $v$, the initial step is filling gap entries with column means:
\begin{equation}
\hat{Y}_{t,h} =
\begin{cases}
Y_{t,h}, & M_{t,h} = 1 \\
\mu_h, & M_{t,h} = 0
\end{cases}
\end{equation}

To standardize the column data, compute scales $\sigma_h=\max\{\mathrm{std}_{\{t:(t,h)\in\Omega\}},\varepsilon\}$ and form
\[
Y = \frac{X^{(0)}-\mu}{\sigma}.
\]

To calculate the low rank projection, compute the singular value decomposition and retain rank-$r$ approximation
\[
Z_r = U_r \Sigma_r V_r^\top.
\]
In which projects the incomplete data onto the leading r principal components.

Finally de-standardize the estimated states:
\[
\hat{Y} = \mu + \sigma \odot Z_r,
\]
and overwrite only on missing entries to obtain $\hat{Y}$.

The iterative step for optimizing the low-rank process can be described as: Standardize $\to$ low-rank projection $\to$ overwrite on missing set until convergence of root-mean-squared-error in missing data fillings. 
\label{Appendix A}

\section{Full comparison of error metrics across ML and traditional methods}
For each variable, let's define $Y_{t,h} \in \mathbb{R}^{d_t \times d_h}$ as ground truth tensor, and $\hat{Y}_{t,h} \in \mathbb{R}^{d_t \times d_h}$ as the reconstructed tensor, and mask tensor as $M \in \{0,1\}^{d_t \times d_h}$.
Over the gaps, Mean Absolute Error (MAE) and Mean Squared Error (MSE) can be defined in Eq. \ref{eq: MAE,MSE}
\begin{equation}
    MAE = \frac{\sum^{d_t}_{t=1} \sum^{d_h}_{h=1}M_{t,h} |Y_{t,h}  - \hat{Y}_{t,h}|}{\sum_{t=1}^{d_t}\sum_{h=1}^{d_h}M_{t,h}} , \qquad
    MSE = \frac{\sum^{d_t}_{t=1} \sum^{d_h}_{h=1}M_{t,h} (Y_{t,h}  - \hat{Y}_{t,h})^2}{\sum_{t=1}^{d_t}\sum_{h=1}^{d_h}M_{t,h}}
    \label{eq: MAE,MSE}
\end{equation}

The Pearson coefficient can be calculated using the Eq. \ref{eq: PR}, and the bar notes taking the mean value over the tensor.
\begin{equation}
    R = \frac{\sum_{t=1}^{d_t}\sum_{h=1}^{d_h}M_{t,h}(Y_{t,h}-\bar{Y})(\hat{Y}_{t,h}-\bar{\hat{Y}})}{\sqrt{\sum_{t=1}^{d_t} \sum_{h=1}^{d_h} M_{t,h} (Y_{t,h} - \bar{Y})^2} \sqrt{\sum_{t=1}^{d_t} \sum_{h=1}^{d_h} M_{t,h} (\hat{Y}_{t,h}-\bar{\hat{Y}})^2}}
    \label{eq: PR}
\end{equation}

Total variation ($TV$), TV difference ($\Delta TV$) and relative TV ($\Delta TV \%$) can be defined in Eq. \ref{eq:TV}:
\begin{align}
    \text{TV} (\mathbf{Y})= \sum_{t=1}^{d_t}\sum_{h=1}^{d_h}M_{t,h}(|Y_{i+1,j} - Y_{t,h}| + |Y_{t,h+1} - Y_{t,h}|) \\ \Delta\text{TV} = \text{TV}(\mathbf{Y}) - \text{TV}(\mathbf{\hat{Y}}), \qquad
    \Delta\text{TV\%} = \frac{\Delta \text{TV}}{\text{TV}(\mathbf{Y})}
    \label{eq:TV}
\end{align}

Table \ref{tbl: error metrics} listed the full error metrics for different variable, which compares the ML approach with other traditional methods mentioned in the methodology subsec. \ref{subsect: Training, Validation, and Test}.

\begin{longtable}{lllSSSS}
\caption{Per-variable error metrics across scenarios (Corr higher is better; all others lower are better).}
\label{tab:pervar_all}\\
\toprule
\textbf{Method} & \textbf{Scenario} & \textbf{Var} &
\textbf{Corr} $\uparrow$ & \textbf{MSE} $\downarrow$ & \textbf{MAE} $\downarrow$ & \textbf{TV} $\downarrow$ \\
\midrule
\endfirsthead

\toprule
\textbf{Method} & \textbf{Scenario} & \textbf{Var} &
\textbf{Corr} $\uparrow$ & \textbf{MSE} $\downarrow$ & \textbf{MAE} $\downarrow$ & $\boldsymbol{\Delta}$\textbf{TV rel} $\downarrow$ \\
\midrule
\endfirsthead

\toprule
\textbf{Method} & \textbf{Scenario} & \textbf{Var} &
\textbf{Corr} $\uparrow$ & \textbf{MSE} $\downarrow$ & \textbf{MAE} $\downarrow$ & $\boldsymbol{\Delta}$\textbf{TV rel} $\downarrow$ \\
\midrule
\endhead

\midrule
\multicolumn{7}{r}{\emph{continued on next page}}\\
\midrule
\endfoot

\bottomrule
\endlastfoot

\multicolumn{7}{l}{\textbf{CT\_MVP}}\\
 & Short  & T & 0.9957 &  12.661247 &  2.540932 & 0.144 \\
 & Short  & U & 0.9872 &  48.454355 &  5.259996 & 0.124 \\
 & Short  & V & 0.9862 &  60.390063 &  5.818168 & 0.125 \\
 & Medium & T & 0.9876 &  35.962065 &  4.165190 & 0.241 \\
 & Medium & U & 0.9622 & 140.214183 &  8.691319 & 0.220 \\
 & Medium & V & 0.9611 & 170.156604 &  9.566741 & 0.224 \\
 & Long   & T & 0.9723 &  90.088569 &  6.473287 & 0.363 \\
 & Long   & U & 0.9267 & 329.916595 & 13.262920 & 0.379 \\
 & Long   & V & 0.9218 & 407.083231 & 14.719980 & 0.389 \\
\hline
\multicolumn{7}{l}{\textbf{Simple\_Transformer}}\\
 & Short  & T & 0.9959 &  11.907560 &  2.486327 & 0.144 \\
 & Short  & U & 0.9848 &  56.233193 &  5.695181 & 0.135 \\
 & Short  & V & 0.9852 &  66.946809 &  6.223970 & 0.146 \\
 & Medium & T & 0.9880 &  34.095705 &  4.083329 & 0.232 \\
 & Medium & U & 0.9571 & 154.535239 &  9.167192 & 0.221 \\
 & Medium & V & 0.9600 & 184.079995 & 10.039750 & 0.247 \\
 & Long   & T & 0.9763 &  79.756331 &  6.438047 & 0.349 \\
 & Long   & U & 0.9174 & 383.512345 & 14.294538 & 0.383 \\
 & Long   & V & 0.9191 & 465.566520 & 15.731714 & 0.412 \\
\hline
\multicolumn{7}{l}{\textbf{PCA}}\\
 & Short  & T & 0.9203 &  216.257420 & 10.101650 & 0.613 \\
 & Short  & U & 0.4197 & 1433.022657 & 28.363168 & 0.882 \\
 & Short  & V & 0.2324 & 1787.432771 & 31.592521 & 0.898 \\
 & Medium & T & 0.9160 &  226.333093 & 10.399497 & 0.640 \\
 & Medium & U & 0.3830 & 1481.448969 & 29.012027 & 0.884 \\
 & Medium & V & 0.2045 & 1817.378905 & 31.886367 & 0.899 \\
 & Long   & T & 0.9121 &  241.225935 & 10.811987 & 0.647 \\
 & Long   & U & 0.3531 & 1613.996485 & 30.241877 & 0.876 \\
 & Long   & V & 0.1582 & 1967.137447 & 33.302031 & 0.890 \\
\hline
\multicolumn{7}{l}{\textbf{Kalman}}\\
 & Short  & T & 0.9104 &  252.557851 & 10.712722 & 0.529 \\
 & Short  & U & 0.3595 & 1523.995653 & 28.476053 & 0.759 \\
 & Short  & V & 0.2008 & 1826.640176 & 31.362302 & 0.789 \\
 & Medium & T & 0.9000 &  273.782092 & 11.083457 & 0.561 \\
 & Medium & U & 0.3328 & 1545.382414 & 28.811210 & 0.798 \\
 & Medium & V & 0.1848 & 1831.748720 & 31.664741 & 0.823 \\
 & Long   & T & 0.8520 &  397.264456 & 12.872844 & 0.557 \\
 & Long   & U & 0.2966 & 1660.271095 & 30.051897 & 0.810 \\
 & Long   & V & 0.1451 & 1924.286747 & 32.584375 & 0.830 \\
\hline
\multicolumn{7}{l}{\textbf{Linear\_Interp}}\\
 & Short  & T & 0.8849 &  362.935649 & 12.453361 & 0.434 \\
 & Short  & U & 0.2564 & 2258.496235 & 33.426859 & 0.538 \\
 & Short  & V & 0.0941 & 2775.124770 & 37.182688 & 0.549 \\
 & Medium & T & 0.8792 &  353.483301 & 12.397414 & 0.495 \\
 & Medium & U & 0.2579 & 2175.635633 & 33.326449 & 0.606 \\
 & Medium & V & 0.1085 & 2750.123323 & 37.536116 & 0.616 \\
 & Long   & T & 0.8761 &  362.879411 & 12.620814 & 0.524 \\
 & Long   & U & 0.2682 & 2194.860185 & 33.705315 & 0.646 \\
 & Long   & V & 0.1148 & 2761.591841 & 37.949697 & 0.654 \\
 \label{tbl: error metrics}
\end{longtable}

\section*{Acknowledgments}
The project is funded by NASA grant 80NSSC24K0124 and NSF grant AGS-2327914. We thank the WACCM-X development team for providing open-access model outputs for machine learning training, and the NCAR Climate Data center for maintaining data availability.

\section*{Author Contributions}

\begin{itemize}[topsep=0em, itemsep=0em, noitemsep,leftmargin=*]
  \item \textbf{Conceptualization:} Jiahui Hu \& Wenjun Dong \& Alan Z. Liu
  \item \textbf{Formal analysis:} Jiahui Hu \& Wenjun Dong \& Alan Z. Liu
  \item \textbf{Investigation:} Jiahui Hu \& Wenjun Dong
  \item \textbf{Software:} Jiahui Hu
  \item \textbf{Supervision:} Wenjun Dong
  \item \textbf{Visualization:} Jiahui Hu \& Wenjun Dong
  \item \textbf{Writing – original draft:} Jiahui Hu 
  \item \textbf{Writing – review \& editing:} Jiahui Hu \& Wenjun Dong \& Alan Z. Liu
  \item \textbf{Funding acquisition:} Wenjun Dong
\end{itemize}

\bibliography{CTMVP}
\bibliographystyle{apacite}

% \section*{Appendix}
% The prediction results for other variables at 90 $km$, i.e. meridional wind ($V$) in Fig. \ref{fig:V_level_idx_60}, temperature ($T$) in Fig. \ref{fig:T_level_idx_60}, and pressure change rage ($\omega$) in Fig. \ref{fig:OMEGA_level_idx_60} are presented in the appendix section. 
% \begin{figure}[!h]
%     \centering
%     \includegraphics[width=\linewidth]{figure/V_level_idx_60.png}
%     \caption{Meridional wind targets versus predictions across different time steps at 90 km.}
%     \label{fig:V_level_idx_60}
% \end{figure}

% \begin{figure}[!t]
%     \centering
%     \includegraphics[width=\linewidth]{figure/T_level_idx_60.png}
%     \caption{Temperature targets versus predictions across different time steps at 90 km.}
%     \label{fig:T_level_idx_60}
% \end{figure}

% \begin{figure}[!b]
%     \centering
%     \includegraphics[width=\linewidth]{figure/OMEGA_level_idx_60.png}
%     \caption{Vertical pressure change rate at 90 $km$, in units of $pa/s$, which equivalent to wind vertical velocity. }
%     \label{fig:OMEGA_level_idx_60}
% \end{figure}
\end{document}